\pdfoutput=1

\documentclass[12pt]{article}
\usepackage[russian, english]{babel}
\usepackage{amsmath}
\usepackage{amsthm}
\usepackage{amssymb}
\usepackage{wrapfig}
\usepackage{amsfonts}
\usepackage{textcomp}
\usepackage{graphicx}
\usepackage{float}
\usepackage{caption}
\usepackage{algpseudocode}
\usepackage{xcolor}
\usepackage{hyperref}

\usepackage{mathenv}
\usepackage{booktabs}
\usepackage{indentfirst}
\usepackage{enumerate}
\usepackage{bbold}
\usepackage{dsfont}
\usepackage{subcaption}

\usepackage{algorithm, algorithmicx, algpseudocode}

\let\le=\leqslant
\let\ge=\geqslant

\let\geq=\geqslant

\newtheorem{Lemm}{Lemma}[section]
\newtheorem{Rem}{Remark}[section]

\newtheorem{Co}{Corollary}[section]
\theoremstyle{definition}

\begin{document}
\baselineskip 12pt

\begin{center}
\textbf{\Large A Method for Generating Connected \\ Erdős–Rényi Random Graphs} \\[1.5ex]
{Boris Chinyaev$^{1}$}\\[0.3cm]
{\small $^{1}$ Lomonosov Moscow State University, bchinyaev.msu@gmail.com}
\end{center}
\vspace{1.0cc}

\begin{abstract}
We propose a novel exact algorithm for generating connected Erd\H{o}s--R\'enyi random graphs $G(n,p)$. The method couples the graph exploration process to an inhomogeneous Poisson random walk, which yields an exact sampler that runs in $O(n)$ time in the sparse regime $p=c/n$. We also show how the method extends to the $G(n,M)$ model via an additional acceptance--rejection step.

\vspace{0.95cc}
\parbox{24cc}{{\it \textbf{Keywords:} random graphs, random walks, generation of connected graphs}}
\end{abstract}

\section{Introduction}

Erdős–Rényi graph models $G(n,p)$ and $G(n,M)$ are fundamental in random graph theory and have numerous applications in network theory, statistical physics, and computer science.
The main probabilistic properties of these models -- including the connectivity threshold -- were first studied in the classic works of Erdős and Rényi (see \cite{ErdosRenyi1960, ErdosRenyi1961}). Connectivity exhibits a sharp threshold at $p=(\log n)/n$: w.h.p connected above, disconnected below; for $p=(\log n+c)/n$ the limiting probability is $e^{-e^{-c}}$.

Many applications require generating a connected random graph.  For instance, in models of communication or social networks, the network structure is often required to be a single connected component. Despite the simplicity of the definitions of the models $G(n,p)$ and $G(n,M)$, generating these graphs conditioned on the connectivity event is a nontrivial task. The main challenge is to sample connected graphs in the sparse regime, which is typical of real-world networks.

A naive approach is to generate a connected graph from $G(n,p)$ via acceptance-rejection sampling. However, such an approach becomes extremely inefficient when the edge probability is small. For example, for $p = c/n$ the probability of connectivity is exponentially small (see \cite{stepanov1970probability, chinyaev2024er_eng}).

Markov chain Monte Carlo (MCMC) methods (see \cite{gray2019generating}) converge to the desired distribution, but it is difficult to analyze the convergence rate. Moreover, they do not yield the exact target distribution in finite time.

Heuristic approaches that generate a random spanning tree and then add edges (see \cite{rodionov2004generating}) fail to produce the correct distribution of $G(n,p)$ conditioned on connectivity. Nevertheless, the two-stage scheme -- first generating a spanning tree -- forms the basis of our algorithm.

In this paper we present an exact algorithm for sampling $G(n,p)$ conditioned on connectivity that runs in polynomial time in the sparse regime $p = c/n$. 
Our construction relies on the representation introduced in \cite{chinyaev2024er_eng}, which expresses the (Markovian) transition law of the graph exploration process in terms of an inhomogeneous Poisson random walk; for background on exploration processes in $G(n,p)$, see \cite{karp1990transitive, nachmias2010critical}. 
This transformation is particularly convenient for studying connectivity in $G(n,p)$ (see also \cite{curien2024erdHos} for a similar perspective) and naturally leads to the proposed sampling procedure: we first generate the exploration tree according to this law and then complete the graph by adding each remaining edge independently with the original probability~$p$.

Furthermore, we demonstrate that our method can be extended to generate connected $G(n,M)$ graphs via an additional acceptance-rejection procedure based on the number of edges.

\section{Preliminary Information}
\subsection{Basic Definitions}

Consider the two classical Erdős–Rényi random graph models.
\begin{enumerate}[\textbullet]
    \item \textbf{The $G(n,p)$ model.} Let $V = \{1,\dots,n\}$ denote the vertex set. Each of the $\binom{n}{2}$ possible undirected edges between pairs of vertices is included in the graph independently with probability $p \in [0,1]$. Then the probability of a given graph $g$ with $m$ edges is
    \begin{equation} \label{Pnp_measure}
    \mathbf{P}_{G(n,p)}(g) = p^{m}(1 - p)^{\binom{n}{2} - m}.
    \end{equation}
    \item \textbf{The $G(n,M)$ model} defines a uniform probability distribution over the set of all undirected graphs with $n$ vertices and exactly $M$ edges. In other words, the probability assigned to any particular graph $g$ with $M$ edges is given by
    \begin{equation} \label{Pnm_measure}
    \mathbf{P}_{G(n,M)}(g) = \binom{\binom{n}{2}}{M}^{-1}, \quad \text{if} \ |E(g)| = M.
    \end{equation}
\end{enumerate}
Let $P_n(p)$ denote the probability that a graph in the $G(n,p)$ model is connected. According to \cite{stepanov1970probability} and \cite{chinyaev2024er_eng}, for $p=c/n$ the following asymptotic relation holds
\begin{equation}\label{Pnp_asymptotics}
P_n(p) =  \left(1- \frac{c \ e^{-c}}{ 1-e^{-c}}\right)\left(1-\left( 1-\frac{c}{n}\right)^{n}\right)^{n} (1+o(1)), \quad n \to \infty.
\end{equation}
This expression shows that the probability of connectivity is exponentially small, which makes a simple \textit{acceptance-rejection} approach impractical.

We are interested in generating a graph under the condition of connectivity, which leads to the following conditional distributions in the corresponding models.
\begin{enumerate}[\textbullet]
    \item In the $G(n,p)$ model, under the connectivity condition, the distribution is given by
    \begin{equation} \label{Pnp_connected}
    \mathbf{P}_{G(n,p)}(g\mid g\text{ is connected}) = \frac{\mathds{1}{\{g\text{ is connected}\}} \  p^{|E(g)|}(1 - p)^{\binom{n}{2} - |E(g)|}}{P_n(p)}.   
    \end{equation}
    \item Similarly, the distribution in the $G(n,M)$ model under the connectivity condition is uniform over the set of all connected graphs with $n$ vertices and $M$ edges:
    \begin{equation}\label{Pnm_connected}
    \mathbf{P}_{G(n,M)}\left( \left. g \right| g\text{ is connected}\right) = \frac{\mathds{1}{\{g\text{ is connected}\}}}{\# \left\{g: |V(g)|=n, \ |E(g)|=M, \ g \text{ is connected}\right\}}.
    \end{equation}
\end{enumerate}

Therefore, the measures given by \eqref{Pnp_connected} and \eqref{Pnm_connected} correspond to the distributions given by \eqref{Pnp_measure} and \eqref{Pnm_measure}, restricted to the subset of connected graphs and normalized accordingly. Since connectivity in the sparse regime is an event of exponentially small probability, the resulting conditional measures differ significantly from the original ones. Despite this, our aim is to construct a practical algorithm capable of generating random connected graphs sampled from these distributions.

\subsection{Exploration Process of $G(n,p)$}
In \cite{chinyaev2024er_eng} we studied the exploration process in the random graph $G(n,p)$. It is used to find the connected component containing a vertex $v$. We briefly review its construction. In this process, vertices can be in one of three states: active, neutral, or explored. Initially, the starting vertex $v_{1}$ is declared active, and all other vertices are neutral:
\begin{equation*}
\mathcal{A}_{1} =\{v_{1} \},\quad \mathcal{U}_{1} =V\setminus \{v_{1} \},\quad \mathcal{R}_{1} =\emptyset .
\end{equation*}
Then, at each step $t$ an active vertex $v_t$ is considered, and all its neutral neighbors become active, while $v_t$ is reclassified as explored:

\begin{equation}\label{AUW}
\begin{gathered}
\mathcal{W}_{t} = \{w: (v_t,w) \in E\},\\[1mm]
\mathcal{A}_{t+1} = \Bigl(\mathcal{A}_{t} \setminus \{v_t\}\Bigr) \cup \mathcal{W}_{t},\quad \mathcal{U}_{t+1} = \mathcal{U}_{t} \setminus \mathcal{W}_{t},\quad \mathcal{R}_{t+1} = \mathcal{R}_{t} \cup \{v_t\}.
\end{gathered}
\end{equation}
The process continues until there are no active vertices left; the final set of explored vertices forms the connected component $\mathcal{C}(v_{1})$. The specific choice of the active vertex at each step is not crucial (for example, one may assume that the first vertex added to the active set is chosen).

Let $A_{t}$ denote the number of the active vertices and $U_{t}$ -- the number of neutral vertices at the beginning of step $t$, and let $W_{t}$ be the number of vertices that become active at this step; the number of the explored vertices coincides with the step number $t$:
\begin{equation*}
A_{t} = |\mathcal{A}_{t}|,\quad U_{t} = |\mathcal{U}_{t}|,\quad W_{t} = |\mathcal{W}_{t}|.
\end{equation*}
We assume that $A_{1} = 1$, $U_{1} = n-1$, hence
\begin{equation*}
A_{t+1} = A_{t} + W_{t} - 1,\quad U_{t+1} = U_{t} - W_{t}.
\end{equation*}
Let $\mathbf{j} = (j_{1}, \dotsc, j_{n})$ denote a trajectory of the process $\{W_{t}\}$.
For the graph to be connected, it is necessary that at each step before $n$ there remains at least one active vertex, i.e., 
$$
A_{t} = 1 + \left(\sum_{\tau=1}^{t} W_{\tau}\right) - t > 0,\quad t < n.
$$ 
Consequently, the connectivity probability can be expressed in terms of this process as follows:
\begin{equation} \label{Pnp1}
\begin{gathered}
P_{n}(p) = \sum_{\mathbf{j} \in J_{n}} \mathbf{P}\Bigl( (W_{1}, \dotsc, W_{n}) = \mathbf{j} \Bigr),\\[1mm]
J_{n} = \left\{ \mathbf{j} : \sum_{i=1}^{k} j_{i} \geq k,\ k < n,\ \sum_{i=1}^{n} j_{i} = n-1 \right\}.
\end{gathered}
\end{equation}
Since the edges in the graph $G(n,p)$ are independent, we get
\begin{equation*}
\mathbf{P}\left( \left. W_{t} = k \ \right| A_{t} = l,\, U_{t} = m \right) = \left\{
\begin{array}{ll}
\binom{m}{k} p^{k} (1-p)^{m-k}, & \text{if } A_{t} > 0,\\[1mm]
0, & \text{if } A_{t} = 0.
\end{array}
\right.
\end{equation*}
Therefore,
\begin{equation}\label{Pnp2}
\mathbf{P}\left( (W_{1}, \dotsc, W_{n}) = \mathbf{j} \right)
= \prod_{t=1}^{n} \left( \binom{n-1 - j_{1} - \dotsb - j_{t-1}}{j_{t}} p^{j_{t}} (1-p)^{j_{t+1} + \dotsb + j_{n}} \right).
\end{equation}

\subsection{Connection with an Inhomogeneous Poisson Random Walk}

In \cite{chinyaev2024er_eng} we had represented expression (\ref{Pnp2}) in the form
\begin{equation}
\mathbf{P}\Bigl( (W_{1}, \dotsc, W_{n}) = \mathbf{j} \Bigr)
= \frac{n! \exp(n)}{n^{n}} \left( 1 - (1-p)^{n} \right)^{n-1} \prod_{t=1}^{n} \left( \exp(-\lambda_{t}) \frac{\lambda_{t}^{j_{t}}}{j_{t}!} \right),
\end{equation}
where
\begin{equation}\label{lambda_i}
\lambda_{i} = \frac{np}{1-(1-p)^{n}} (1-p)^{(i-1)}, \quad \sum_{i=1}^{n} \lambda_{i} = n.
\end{equation}
Hence, the following relation holds:
\begin{equation}\label{PW=j}
\mathbf{P}\Bigl( (W_{1}, \dotsc, W_{n}) = \mathbf{j} \Bigr)
= \frac{n! \exp(n)}{n^{n}} \left( 1 - (1-p)^{n} \right)^{n-1} \ \mathbf{P}\Bigl( (X_{1}, \dotsc, X_{n}) = \mathbf{j} \Bigr),
\end{equation}
where $X_{i}$ are independent random variables with $X_{i} \sim \mathrm{Poiss}(\lambda_{i})$.
Then, summing expressions (\ref{PW=j}) by $\mathbf{j} \in J_{n}$, we get
\begin{equation}\label{PnpJX}
\begin{gathered}
P_{n}(p) = \mathbf{P}\Bigl( (W_{1}, \dotsc, W_{n}) \in J_{n} \Bigr) = \\
= \mathbf{P}\Bigl(\sum_{i=1}^{n} X_{i} = n-1\Bigr)^{-1} \left( 1 - (1-p)^{n} \right)^{n-1} \mathbf{P}\Bigl( (X_{1}, \dotsc, X_{n}) \in J_{n} \Bigr).
\end{gathered}
\end{equation}
Thus, we obtain the following lemma.
\begin{Lemm}[\cite{chinyaev2024er_eng}]\label{connect}
Let $G \sim G(n,p)$ be an Erdős–Rényi graph. Then the connectivity probability of $G$ is given by
$$
P_n(p) = \left( 1 - (1-p)^{n} \right)^{n-1} \mathbf{P}\left( \left.S_k \geq 0,\ 0 < k < n \right| S_{n} = -1 \right),
$$
where $S_k = \sum_{i=1}^{k} (X_i - 1)$, and the $X_i$ are independent random variables $X_i \sim \mathrm{Poiss}(\lambda_i)$. Here $\lambda_i$ are defined in (\ref{lambda_i}).
\end{Lemm}

From the above, it follows that one can obtain the distribution of the trajectory of the process $\{W_{k}\}$ in a connected graph. More precisely, we can deduce how it is expressed in terms of the trajectory distribution of the random walk $\{X_{k}\}$, dividing expression (\ref{PW=j}) on (\ref{PnpJX}). Thus, we derive the following corollary, obtaining conditional distributions of the trajectories.

\begin{Co}\label{connect_co}
Under the conditions of Lemma \ref{connect} the following relations hold:
\begin{equation}
\begin{gathered}
\mathbf{P}\left( \left.(W_{1}, \dotsc, W_{n}) = \mathbf{j} \ \right| G\text{ is connected} \right) = \\
= \mathbf{P}\left( \left.(W_{1}, \dotsc, W_{n}) = \mathbf{j} \ \right|(W_{1}, \dotsc, W_{n}) \in J_{n} \right) = \\
= \mathbf{P}\left( \left.(X_{1}, \dotsc, X_{n}) = \mathbf{j} \ \right| (X_{1}, \dotsc, X_{n}) \in J_{n} \right).
\end{gathered}
\end{equation}
\end{Co}

This corollary is crucial for an algorithm for generating connected graphs. Since there is an equality between the distributions of the processes $\{W_{k}\}$ and $\{X_{k}\}$ on the conditional spaces, we can generate their trajectories using the space of Poisson random variables $\{X_{k}\}$.

In this space, in the case $p=c/n$, the event that a trajectory remains non-negative occurs with a certain positive probability (asymptotically independent of $n$). We discuss this probability in Section \ref{Complexity1} analyzing the algorithm's complexity.

\section{Generation of Connected $G(n,p)$}

Based on the discussion above, we propose the following generation scheme. First, we generate the exploration trajectory of the graph. Then, given a fixed exploration, we construct the graph. The overall procedure is described in Algorithm~\ref{alg:connected-gnp}. Below, we provide a detailed proof of the correctness of this procedure.

\begin{algorithm}[h]
\caption{Generation of a connected $G(n,p)$ graph}
\label{alg:connected-gnp}
\begin{algorithmic}[1]
\Require Number of vertices $n \ge 1$, edge probability $0 < p \le 1$
\Ensure Connected graph $G$ drawn from $G(n,p)$ 

\vspace{0.3em}
\Statex \textbf{Step 1. Generate the exploration trajectory}
\vspace{0.3em}
\State $\displaystyle \lambda_i \gets \frac{np}{1 - (1 - p)^n} (1 - p)^{i - 1}$ for $i = 1, \dots, n$
\vspace{0.3em}
\Repeat
    \State Generate $(X_1,\dots,X_n) \sim \mathrm{Multinomial}\!\bigl(n-1;\, \tfrac{\lambda_1}{n}, \dots, \tfrac{\lambda_n}{n}\bigr)$
    \State $S_k \gets \sum_{i=1}^{k} (X_i - 1)$ for $k = 1,\dots,n$
\Until{$S_k \ge 0$ for all $k < n$}
\vspace{0.3em}
\Statex \textbf{Step 2. Construct the exploration tree}
\State $\mathcal{A}_1 \gets \{v_1\}$,\quad $\mathcal{U}_1 \gets V \setminus \{v_1\}$
\For{$t = 1$ to $n$}
    \State Select $v_t \in \mathcal{A}_t$
    \State Sample uniformly $\mathcal{W}_t \subset \mathcal{U}_t$ with $|\mathcal{W}_t| = X_t$ 
    \Comment{$\Pr = \binom{|\mathcal{U}_t|}{X_t}^{-1}$}
    \ForAll{$w \in \mathcal{W}_t$}
        \State Add edge $(v_t, w)$ to $E(G)$
    \EndFor
    \State Update $\mathcal{A}_{t+1}, \mathcal{U}_{t+1}$ via \eqref{AUW}
\EndFor
\vspace{0.3em}

\Statex \textbf{Step 3. Add the remaining edges}
\State $\mathcal{P} \gets \bigcup_{t} \bigl\{(v_t,w) : w \in \mathcal{A}_t \setminus \{v_t\}\bigr\}$
\ForAll{$(u,v) \in \mathcal{P}$}
    \State Insert edge $(u,v)$ into $E(G)$ with probability $p$
\EndFor
\State \Return $G$
\end{algorithmic}
\end{algorithm}

\subsection{Generation of the Exploration}
According to Corollary \ref{connect_co}, under the condition that a $G(n,p)$ graph is connected, the distribution of the sequence $\{W_k\}$ coincides with the distribution of $\{X_k\}$, where $X_k$ are independent random variables with $X_k \sim \mathrm{Poiss}(\lambda_k)$, conditioned on
\begin{equation}\label{stop_condition}
S_n = \sum_{t=1}^n (X_t-1) = - 1,
\quad
S_k = \sum_{t=1}^k (X_t - 1) \ge 0,\quad k < n.
\end{equation}
A natural way to sample such trajectories is an acceptance--rejection scheme: we repeatedly draw $X_{1},\dots,X_{n}$ and keep the first realization that satisfies the condition \eqref{stop_condition}.

\begin{figure}[h]
    \centering
    \includegraphics[width=0.8\linewidth, height=0.4\linewidth]{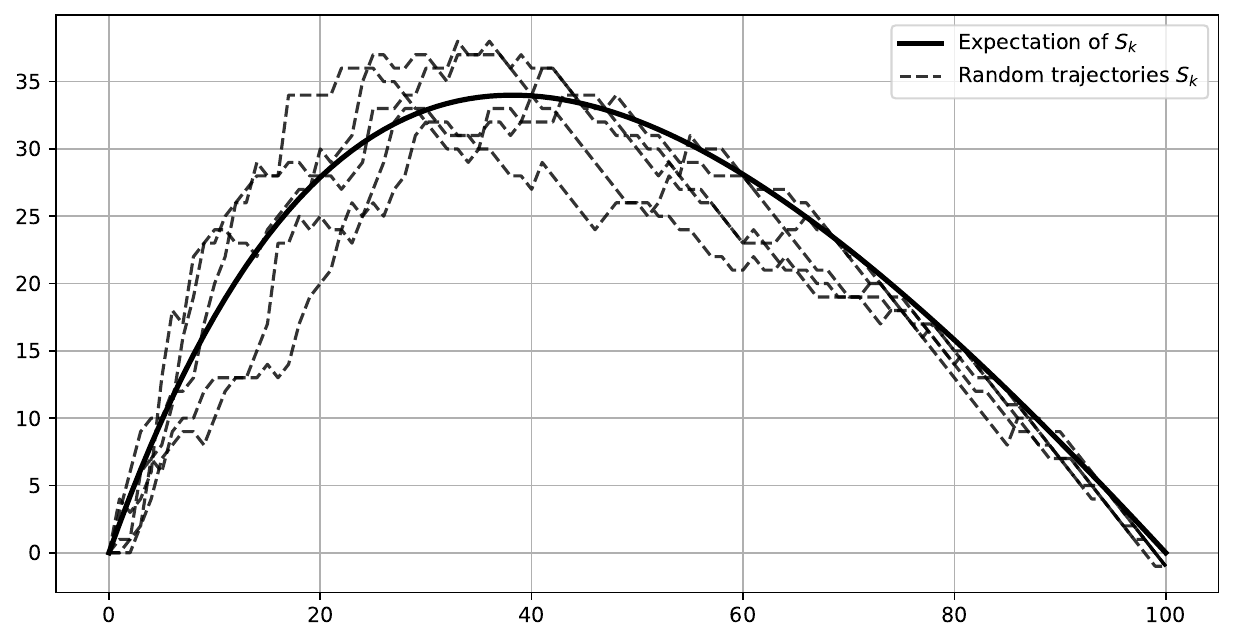}  
    \captionsetup{justification=centering} 
    \caption{Plot of the expected value and examples \\ of realizations of $S_k$ for $n=100$, $p = 3/n$.}
    \label{ex1}
\end{figure}

The trajectory generation procedure described above is already performs well in practical settings. Nevertheless, it can be further accelerated by restricting the sampling to only those trajectories that satisfy the condition $S_n = -1$. This can be achieved by using the following identity in distribution:
\begin{equation*}
\begin{gathered}
\mathbf{P} \left( (X_{1}, \dotsc, X_{n}) = \mathbf{j} \ \left| \ \sum_{t=1}^{n} X_t = n - 1 \right. \right)
= \mathbf{P} \left( (Y_{1}, \dotsc, Y_{n}) = \mathbf{j} \right), \\
(Y_{1}, \dotsc, Y_{n}) \sim \mathrm{Multinomial}\left(n - 1;\, \tfrac{\lambda_1}{n}, \dotsc, \tfrac{\lambda_n}{n} \right).
\end{gathered}
\end{equation*}
This multinomial formulation is used in Step~1 of Algorithm~\ref{alg:connected-gnp}.
Examples of random walks $S_k$ obtained by this method are shown in Figure~\ref{ex1}.

\subsection{Generation of a Graph for a Fixed Exploration}
Now, consider the probability to obtain a specific graph $g$ given a fixed trajectory of the process $\{W_k\}$, that is, the distribution
\begin{equation*}
\mathbf{P}\Bigl( G = g \ \Bigl| \ (W_1,\dots,W_n) = \mathbf{j} \Bigr).
\end{equation*}
Next, we will show that this measure coincides with the measure obtained by the following procedure (Steps 2 and 3 of Algorithm~\ref{alg:connected-gnp}).

\paragraph{Reconstruction of the Tree from the Trajectory.} Let the exploration trajectory of the graph be given by $\mathbf{j} = (j_1, j_2, \dots, j_n)$. We construct an exploration tree that is consistent with this trajectory. Considering the steps $t = 1, \dots, n$, we build the sets $\mathcal{A}_t$ and $\mathcal{U}_t$ according to (\ref{AUW}). The set $\mathcal{W}_t$ is constructed as follows:
\begin{enumerate}[1.]
    \item At step $t$, an active vertex $v_t \in \mathcal{A}_t$ is selected according to a fixed selection rule.
    \item Then, from the current set of unexplored vertices $\mathcal{U}_t$, a subset $\mathcal{W}_t$ of size $j_t$ is selected uniformly at random, i.e., every subset $\mathcal{W}_t \subseteq \mathcal{U}_t$ with $|\mathcal{W}_t| = j_t$ is chosen with probability $\binom{|\mathcal{U}_t|}{j_t} ^ {-1}$.
    \item For each vertex $w \in \mathcal{W}_t$, the edge $(v_t, w)$ is added to the exploration tree. These edges form the set $\mathcal{T}$.
\end{enumerate}
\newpage

Once the exploration tree is finished, every pair of vertices can be classified into one of the following three disjoint sets:

\begin{enumerate}[\textbullet]
\item $\mathcal{T}$: pairs of vertices connected by the edges of the constructed exploration tree. These are exactly the edges used to discover new vertices during the exploration:
\begin{equation*}
\mathcal{T} = \bigcup_{t} \{ (v_t, w) : w \in \mathcal{W}_t \};
\end{equation*}

\item $\mathcal{F}$: pairs $(v, w)$ such that, at the moment when $v$ was selected as an active vertex, $w$ was still unexplored (i.e., in the neutral set $\mathcal{U}_t$), but the edge $(v, w)$ was not selected to be part of the exploration tree:
\begin{equation*}
\mathcal{F} = \bigcup_{t} \{ (v_t, w) : w \in \mathcal{U}_t \setminus \mathcal{W}_t \};
\end{equation*}

\item $\mathcal{P}$: all remaining unordered pairs of vertices, which were not considered during the exploration process. The presence of edges between these pairs has not yet been determined:
\begin{equation*}
\mathcal{P} = \{ (v, w) : (v, w) \notin \mathcal{T},\ (v, w) \notin \mathcal{F} \}
= \bigcup_{t} \{ (v_t, w) : w \in \mathcal{A}_t \setminus \{v_t\} \}.
\end{equation*}
\end{enumerate}

\paragraph{Generation of the Remaining Edges.}
Thus, we have constructed the exploration tree and partitioned all pairs of vertices into the disjoint sets $\mathcal{T}$, $\mathcal{F}$, and $\mathcal{P}$. Then, for all pairs belonging to the set $\mathcal{P}$, each edge is sampled independently with inclusion probability $p$. Therefore, the set of edges of the final graph $g$ is given by
$$
E(g) = \mathcal{T} \cup \{\, (u,v) \in \mathcal{P} : \xi_{u,v} = 1 \,\},
$$
where $\xi_{u,v} \sim \mathrm{Bernoulli}(p)$ are independent for all $(u,v) \in \mathcal{P}$.

\subsection{Explanation of the Method's Correctness} The correctness of this procedure is based on the following reasoning. Indeed, the sequence of sets $\mathbf{w} = (\mathcal{W}_1, \dots, \mathcal{W}_n)$ determines only the presence in the graph $g$ of edges from the set $\mathcal{T}$ and the absence of edges from the set $\mathcal{F}$. Formally, the following relation holds:
\begin{equation*}
\left\{ g : (\mathcal{W}_1, \dots, \mathcal{W}_n) = \mathbf{w} \right\} \Leftrightarrow \left\{ g : \left\{ \bigcap_{e \in \mathcal{T}(\mathbf{w})} e \in E(g) \right\} \cap \left\{ \bigcap_{e \in \mathcal{F}(\mathbf{w})} e \notin E(g) \right\} \right\}.
\end{equation*}
This relation holds in one direction by construction. Moreover, $\mathbf{w}$ can be uniquely recovered from $\mathcal{T}(\mathbf{w})$ and $\mathcal{F}(\mathbf{w})$. Hence, due to the overall independence of the edges in the $G(n,p)$ model, we obtain
\begin{equation*}
\mathbf{P}\Bigl( (\mathcal{W}_1, \dots, \mathcal{W}_n) = \mathbf{w} \Bigr) = p^{|\mathcal{T}(\mathbf{w})|} (1-p)^{|\mathcal{F}(\mathbf{w})|}.
\end{equation*}
Similarly, by the definition of the $G(n,p)$ model, for the remaining pairs of vertices (from the set $\mathcal{P}$) the edge states are determined by independent trials with probability $p$. Therefore, we obtain
\begin{gather*}
\mathbf{P}\Bigl( G = g \ \Bigl| \ (\mathcal{W}_1, \dots, \mathcal{W}_n) = \mathbf{w} \Bigr) = \\
\prod_{e \in \mathcal{T}(\mathbf{w})} \mathds{1}\{e \in g\} \; \prod_{e \in \mathcal{F}(\mathbf{w})} \mathds{1}\{e \notin g\} \; \prod_{e \in \mathcal{P}(\mathbf{w})} p^{\mathds{1}\{e \in g\}} (1-p)^{\mathds{1}\{e \notin g\}}.
\end{gather*}

In our procedure, at step $t$, a subset of $j_t$ vertices is selected uniformly at random without replacement from the set $\mathcal{U}_t$ (with probability $1/\binom{|\mathcal{U}_t|}{j_t}$). This is consistent with the desired distribution due to the symmetry of the model. Indeed, $|\mathcal{T}(\mathbf{w})|$ and $|\mathcal{F}(\mathbf{w})|$ depend only on $|\mathbf{w}| = (|\mathcal{W}_1|, \dots, |\mathcal{W}_n|)$:
\begin{equation*}
|\mathcal{T}(\mathbf{w})| = \sum_{t=1}^{n} |\mathcal{W}_t|,\quad |\mathcal{F}(\mathbf{w})| = \sum_{t=1}^{n} \left( n-1 - \sum_{\tau=1}^{t} |\mathcal{W}_{\tau}| \right).
\end{equation*}
Hence,
\begin{gather*}
\mathbf{P}\Bigl( (\mathcal{W}_1, \dots, \mathcal{W}_n) = \mathbf{w} \ \Bigl| \ (W_1, \dots, W_n) = \mathbf{j} \Bigr) \\
= \mathds{1}\{|\mathbf{w}| = \mathbf{j}\} \frac{p^{|\mathcal{T}(\mathbf{w})|} (1-p)^{|\mathcal{F}(\mathbf{w})|}}{\sum\limits_{\tilde{\mathbf{w}} :\, |\tilde{\mathbf{w}}| = \mathbf{j}} p^{|\mathcal{T}(\tilde{\mathbf{w}})|} (1-p)^{|\mathcal{F}(\tilde{\mathbf{w}})|} } 
= \mathds{1}\{|\mathbf{w}| = \mathbf{j}\} \frac{j_1! \dots j_n!}{n!}.
\end{gather*}
This indeed corresponds to our procedure.

\subsection{Complexity Analysis}\label{Complexity1}
We evaluate the step-by-step complexity of Algorithm~\ref{alg:connected-gnp} with primary focus on the sparse case $p=c/n$ (fixed $c>0$), while also addressing $p$ below and above this scale.

\begin{enumerate}[1.]
\item \emph{Generation of the exploration trajectory (Step~1).} The acceptance--rejection scheme requires repeated sampling of a multinomial vector \((X_1,\dots,X_n)\) until the random walk $\{S_k\}$ stays nonnegative.
According to \cite{chinyaev2024er_eng}, two key estimates hold for the probability that the walk remains nonnegative.
\begin{enumerate}[\textbullet]
    \item \emph{Non-asymptotic lower bound}:
    $$
    \mathbf{P}\left(\left. S_k \geq 0, \ 0 < k < n \ \right| S_{n} = -1 \right) \ge \frac{1}{n},
    $$
    so even for very small \(p\) the expected number of restarts is \(O(n)\).
    \item \emph{Asymptotic estimate} as $n \to \infty$, $p = c/n$:
    \begin{equation}\label{positivity_asympt}
    \mathbf{P}\left(\left. S_k \geq 0, \ 0< k < n \ \right| S_{n} = -1 \right)
    \to
    \bigl(1 - e^{-c}\bigr)\,\bigl(1 - \tfrac{c\,e^{-c}}{1 - e^{-c}} \bigr).
    \end{equation}
    In this case, the acceptance probability tends to a positive constant, and the expected number of restarts is $O(1)$.
\end{enumerate}
Sampling a multinomial vector of length $n$ costs $O(n)$.  
Hence, Step~1 runs in $O(n^2)$ time in the general case, but in the regime $p=c/n$ it is $O(n)$ on average.

\item \emph{Construction of the exploration tree (Step~2).} 
The naive implementation draws $X_t$ vertices uniformly from $\mathcal U_t$ at every step.
Equivalently, one can generate a single random permutation $\sigma$ of $[n]$ and slice it according to the block sizes \(X_t\).
This makes Step~2 linear: \(O(n)\).

\item \emph{Addition of the remaining edges (Step~3).} 
\begin{enumerate}[\textbullet]
    \item The number of unchecked vertex pairs (in the set $\mathcal{P}$) is of the order $O(n^2)$. Testing each candidate edge independently with probability $p$ takes $O(n^2)$ time.
    \item In the sparse case $p=c/n$, we avoid scanning all $O(n^2)$ pairs by sampling the \emph{geometric waiting times} between successes (see \cite{batagelj2005efficient}).
Since the expected number of additional edges is $p\binom{n}{2}=O(n)$, this yields the successful indices in expected $O(n)$ time. Moreover, knowing the discovery order $\sigma$ and the walk $(S_t)$, and using the following representation
$$
\mathcal{P} = \bigcup_{t} \{ (\sigma(t), \sigma(w)): w \in \{t+1, \dots, t + S_{t-1} \} \}, 
$$
we can map these indices to vertex pairs in overall $O(n)$ time.

\end{enumerate}
\end{enumerate}

\noindent\textbf{Summary.}
For $p=c/n$, each step is $O(n)$ in expectation (by \eqref{positivity_asympt} and geometric skipping in Step~3), so the overall running time is expected $O(n)$.
For smaller $p$, the acceptance–rejection in Step~1 may dominate, but the total cost remains at most $O(n^2)$ by the nonasymptotic bound above.
For larger $p$, Step~3 dominates; with geometric skipping the expected cost scales linearly with $m := p\binom{n}{2}$, hence $O(n^2)$ in the worst case.

\begin{Rem}
The refinements discussed in this section are not shown explicitly in Algorithm~\ref{alg:connected-gnp}.
However, we have included the equivalent permutation-based implementation via $\sigma$ in Algorithm~\ref{alg:connected-gnm}.
\end{Rem}

\subsection{Experimental Results}

In this section, we present visualizations of graphs generated using the proposed algorithm, as well as empirical observations that confirm the conformity with the desired distribution.

\begin{figure}[H]
    \centering
    \includegraphics[width=0.9\linewidth, height=0.5\linewidth]{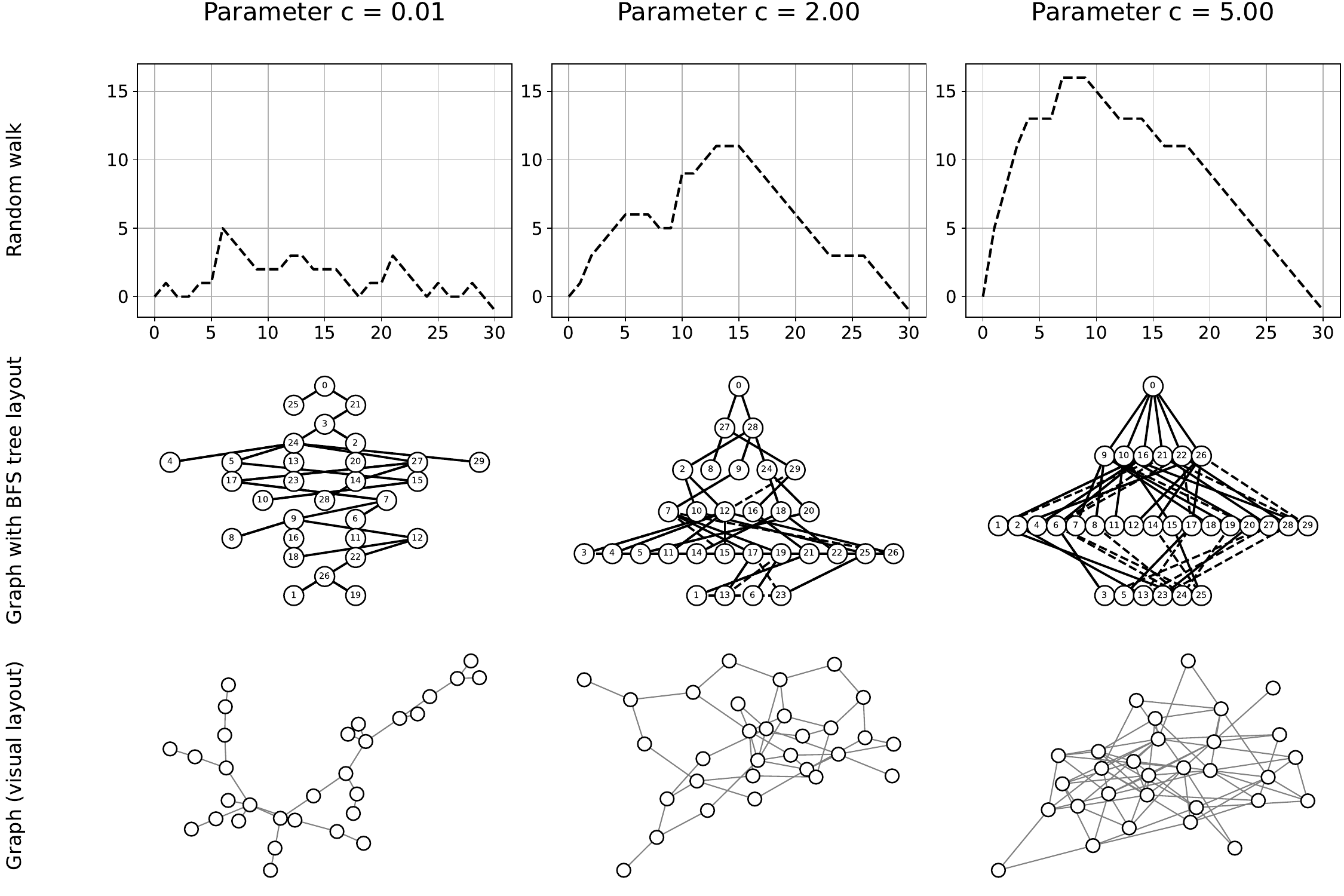}  
    \caption{Visualization of random graphs for $n=30$ and different values of $c$.}
    \label{fig:visualization}
\end{figure}

Figure~\ref{fig:visualization} shows examples of generating graphs $G(n,p)$ for $p = c/n$ for various values of $c$. For each case, the following are displayed: the trajectory of the random walk $\{S_k\}$, the constructed exploration tree, and the final connected graph.

\newpage

\begin{figure}[H]
    \centering
\includegraphics[width=0.95\linewidth]{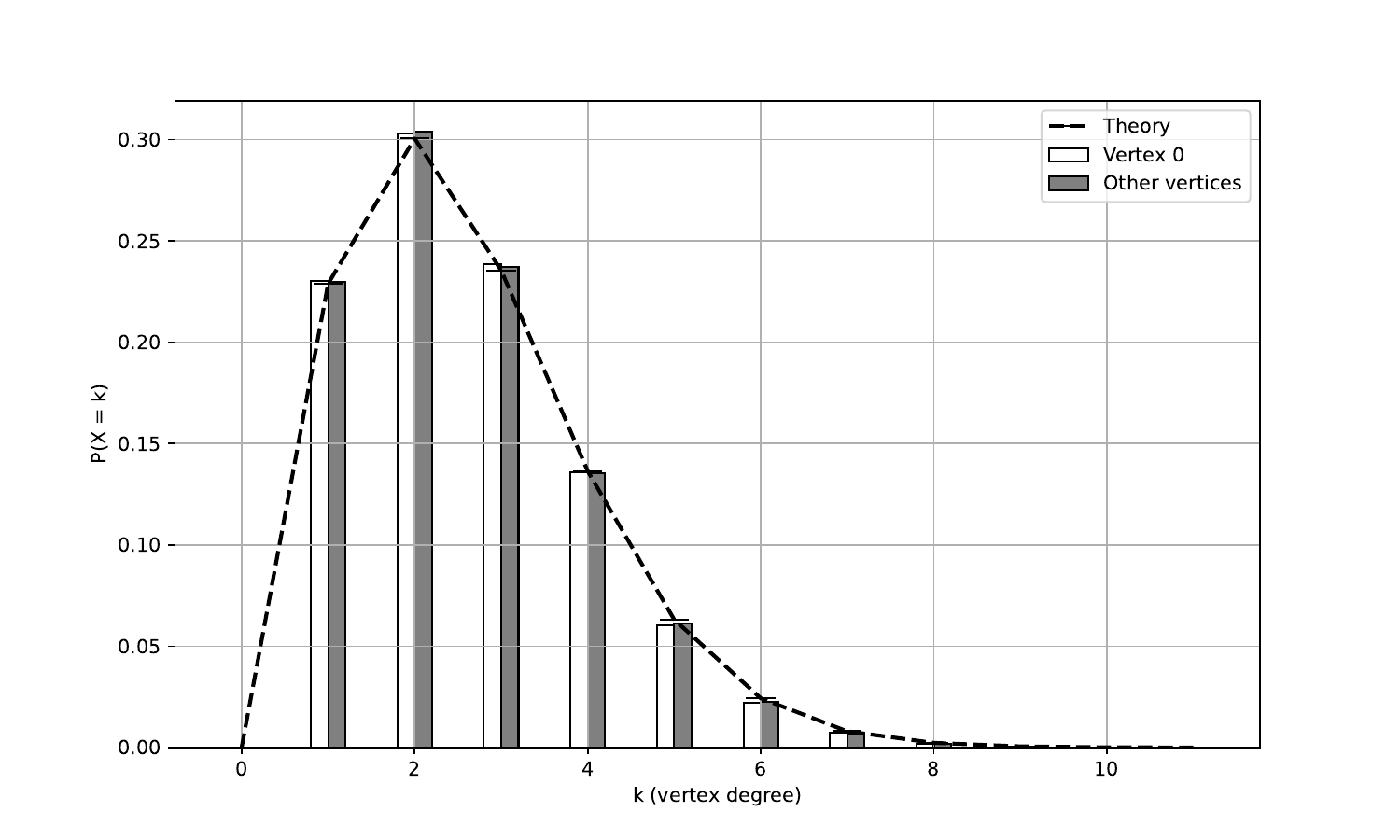}
    \caption{Empirical vertex degree distribution for  $c = 2$, $n = 100$.}
    \label{fig:degree-vs-index}
\end{figure}

Despite the asymmetry of the generation procedure, the final distribution of graphs turns out to be symmetric. In particular, distribution of a vertex degree does not depend on its index, as seen in Figure~\ref{fig:degree-vs-index}. The figure also shows the theoretical vertex degree distribution discussed in Remark \ref{zeta_remark}.

\begin{Rem}\label{zeta_remark}
The number of neighbors of a vertex in a connected graph corresponds to the distribution of a random variable $X_1$ conditioned on the exploration walk $\{S_k\}$ staying nonnegative. Its asymptotic law is well approximated by
\[
\mathbf{P}(X_1 = k) \approx e^{-\gamma}\,\frac{\gamma^{k}}{k!}\,\frac{1-e^{-ck}}{1-e^{-c}}, \quad k>0,\quad \gamma=\frac{c}{1-e^{-c}}.
\]
A rigorous proof of this statement is nontrivial and is omitted here; however, it is clearly supported by our experiments. In particular, it implies the following expression for the average degree of a vertex in a connected graph (which coincides with the expression obtained in \cite[§4]{bender1992asymptotic}):
\begin{equation}\label{zeta}
\begin{gathered}
\zeta(c)=\sum_{k=1}^{\infty} k\,e^{-\gamma}\frac{\gamma^{k}}{k!}\frac{1-e^{-ck}}{1-e^{-c}}
=\frac{\gamma - \gamma e^{-c}\exp\!\bigl(\gamma(e^{-c}-1)\bigr)}{1-e^{-c}}
=c\,\frac{1+e^{-c}}{1-e^{-c}}=\frac{c}{\tanh(c/2)}.
\end{gathered}
\end{equation}
This expression is used in the generation procedure for connected $G(n,M)$ graphs.
\end{Rem}

\begin{figure}[H]
    \centering
        \includegraphics[width=0.95\linewidth, height=0.6\linewidth]{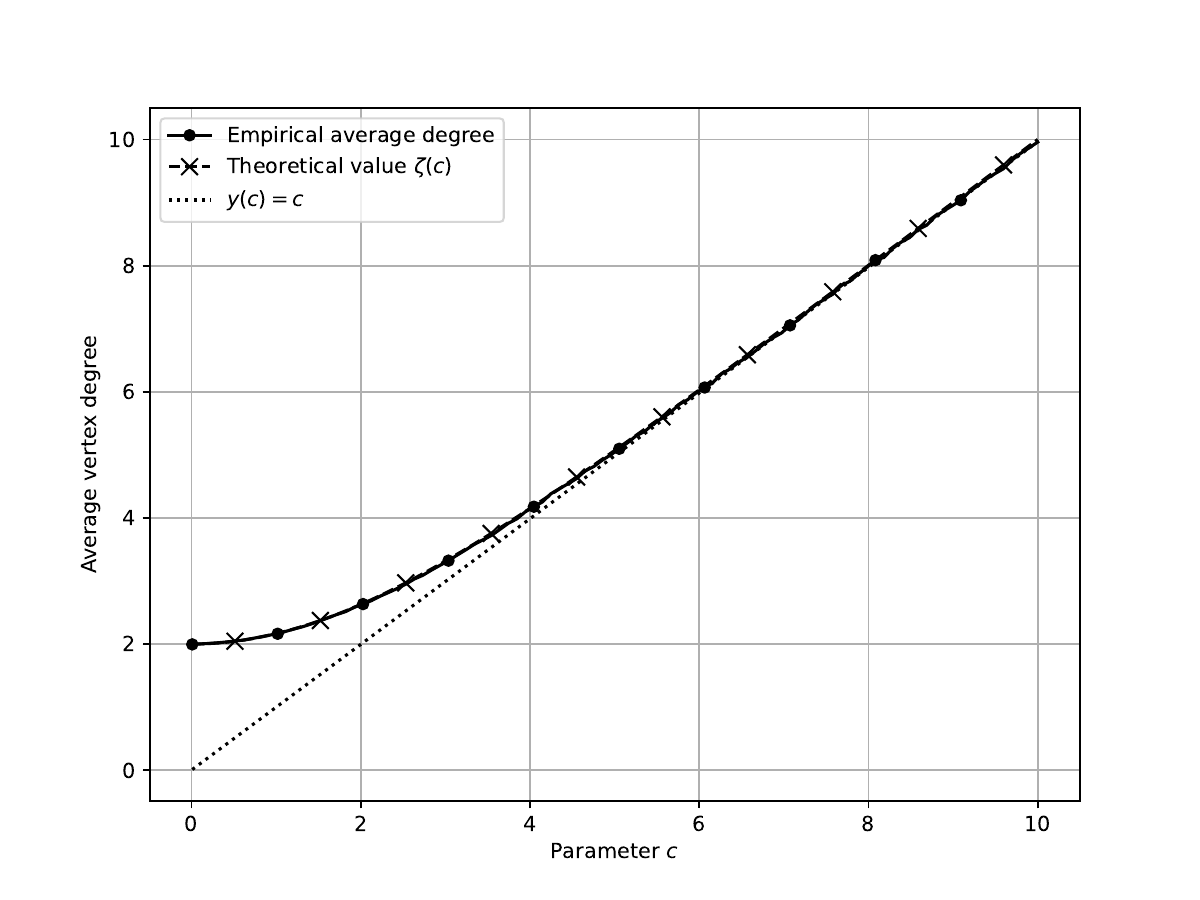}
    \caption{Average vertex degree as a function of parameter $c$ for $n = 300$.}
    \label{fig:degree-vs-c}
\end{figure}

Figure~\ref{fig:degree-vs-c} shows the dependence of the empirical and theoretical (given by (\ref{zeta})) average vertex degree on the parameter $c$ for a fixed $n$. As $c \to 0$, it tends to $2$ (which corresponds to a tree), and for large $c$ it approaches $c$ — as in the unconstrained $G(n,p)$ model.

\section{Generation of Connected Graphs $G(n,M)$}

\subsection{Generation Algorithm idea}\label{section:alg2}

In this section, we derive a method for generation of connected $G(n,M)$ graphs. As in the case of $G(n,p)$, we are interested in exact generation, that is, a method that yields a uniform distribution over the set of all connected graphs with $n$ vertices and $M$ edges.

The method is based on the following idea: we use the already constructed exact algorithm for generating connected graphs $G(n,p)$, choosing the parameter $p$ appropriately. Then, we apply an acceptance-rejection procedure, accepting only those graphs that have exactly $M$ edges. The overall procedure is described in Algorithm~\ref{alg:connected-gnm}.

\begin{algorithm}[h]
\caption{Generation of a connected $G(n,M)$ graph}
\label{alg:connected-gnm}
\begin{algorithmic}[1]
\Require Number of vertices $n \ge 1$, number of edges $n-1 \le M \le \binom{n}{2}$
\Ensure Connected graph $G$ drawn from $G(n,M)$

\Statex \textbf{Step 0. Compute $p$}
\State Choose $p = c/n$ such that $\zeta(c) = 2M/(n-1)$ (see \eqref{zeta})
\vspace{0.3em}
\Statex \textbf{Step 1. Generate the exploration trajectory}
\vspace{0.3em}
\State $\displaystyle \lambda_i \gets \frac{np}{1 - (1 - p)^n} (1 - p)^{i - 1}$ for $i = 1, \dots, n$
\Repeat
    \State Generate $(X_1,\dots,X_n) \sim \mathrm{Multinomial}\!\bigl(n-1;\, \tfrac{\lambda_1}{n}, \dots, \tfrac{\lambda_n}{n}\bigr)$
    \State $S_k \gets \sum_{i=1}^{k} (X_i - 1)$ for $k=1,\ldots,n$
    \State Draw $E_p \sim \mathrm{Binomial}\!\bigl(\sum_{i=1}^{n-1} S_i, p\bigr)$
\Until{$S_k \ge 0$ for all $k < n$ and $E_p = M-(n-1)$}
\vspace{0.3em}

\Statex \textbf{Step 2. Construct the exploration tree}
\State $k \gets 1$,  $\sigma \gets$ uniform random permutation of $[n]$
\For{$t = 1$ to $n$}

    \For{$w = k+1$ to $k+X_t$}
        \State Add edge $(\sigma(t), \sigma(w))$ to $E(G)$
    \EndFor
    \State $k \gets k + X_t$ 
\EndFor
\vspace{0.3em}

\Statex \textbf{Step 3. Add the remaining edges}
\State Select $\{i_1, \dots, i_{M-(n-1)}\}$ uniformly from $\left\{1,\dots, \sum_{i=1}^{n-1} S_i \right\}$ 
\State $i \gets 0$, $t \gets 1$ 
\For{$j = 1$ to $M-(n-1)$}
    \While {$i + S_{t-1}< i_j$}
        \State $i \gets i + S_{t-1}$, $t \gets t + 1$ 
    \EndWhile
    \State Add edge $(\sigma(t), \sigma(t+i_j-i))$ to $E(G)$
\EndFor
\State \Return $G$
\end{algorithmic}
\end{algorithm}

\subsection{Explanation of the Method's Correctness}

\begin{figure}[H]
    \centering
    \includegraphics[width=0.9\linewidth, height=0.45\linewidth]{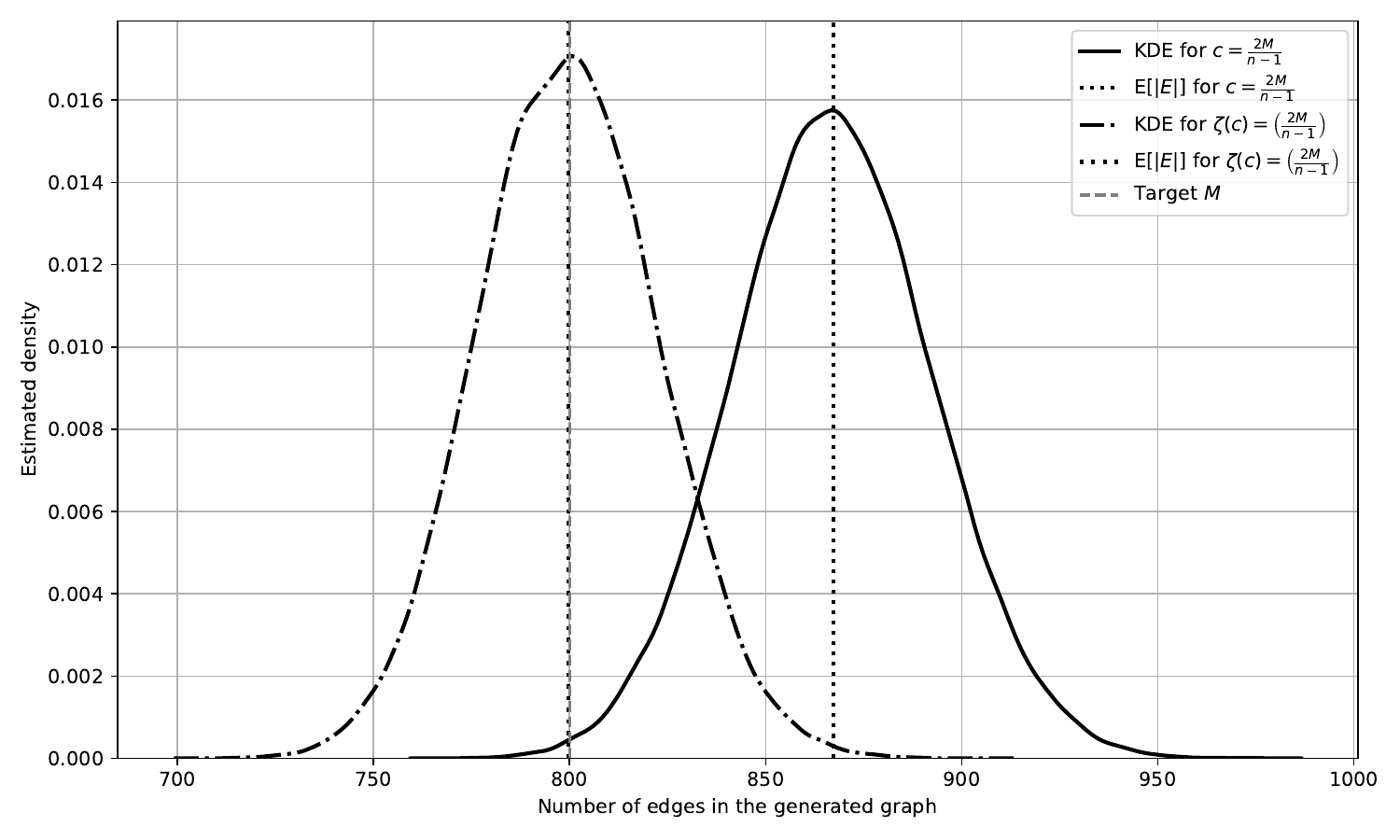}
    \caption{Distribution of the number of edges when generating $G(n,M)$ via $G(n,p)$}
    \label{fig:gnm-ru}
\end{figure}

Consider a graph $g \sim G(n,p)$ for any fixed $p$. In this model, all graphs with $M$ edges have the same probability:
\[
\mathbf{P}_{G(n,p)}(g) = p^M (1 - p)^{\binom{n}{2} - M},\quad \text{if } |E(g)| = M.
\]
Thus, the conditional distribution
$$
\mathbf{P}_{G(n,p)}\left(g \ \bigl| \ g\text{ is connected},\ |E(g)| = M \right)
$$
is uniform over the set of all connected graphs with $n$ vertices and $M$ edges. This is exactly the distribution of connected graphs in the $G(n,M)$ model. Hence, if we generate connected $G(n,p)$ graphs using Algorithm~\ref{alg:connected-gnp} until $|E(G)| = M$, we obtain the correct distribution. However, this scheme is inefficient without an additional optimization.
\begin{enumerate}[1)]
    \item We immediately (in Step 1) generate the number of edges $E_p$ that will be added in Step 3, since it is known that
    $$\displaystyle E_p \sim  \mathrm{Binomial}\! \left(\sum_{i=1}^{n-1} S_i, p\right).
    $$
    In this way, we only repeat the operations of the first step, which in the typical case ($p=c/n$) has a complexity of $O(n)$. In Step 3, we then choose the already known number of edges uniformly from the set $\mathcal{P}$.
    \item For the algorithm to work correctly, we must reject all trajectories $\{S_k\}$ if the desired value of $E_p$ is not obtained. Therefore, we must reduce the number of regenerations by choosing an optimal $p$. It is proposed to choose the parameter $p = c/n$ such that $\zeta(c) = 2M/(n-1)$ (see Remark \ref{zeta_remark}). A~comparison of the random $M$ obtained by this approach and by the naive approach $c = 2M/(n-1)$ is shown in Figure \ref{fig:gnm-ru}.
\end{enumerate}

\subsection{Remarks on complexity}

The $G(n,M)$ algorithm differs from the $G(n,p)$ one only minimally in Steps~2 and~3. 
These steps are essentially unchanged: Step~2 is $O(n)$, and Step~3 is linear in the number of extra edges $m:=M-(n-1)$.

Step~0 solves $\zeta(c)=2M/(n-1)$ for $c$ and sets $p=c/n$; we treat this inversion as $O(1)$ (e.g., a constant number of Newton steps), since $\zeta$ is smooth and monotone.

The delicate part is Step~1. By enforcing the positivity condition ($S_k \ge 0$ for $k < n$), Step~1 ensures that the resulting graph is connected. Let $X_p(n)$ denote the number of edges in $G(n,p)$ under the conditional law given that the graph is connected. Then the remaining acceptance event $E_p=M-(n-1)$ is equivalent to $X_p(n)=M$. Here we can apply results from \cite[§4]{bender1992asymptotic}. By choosing $p=c/n$ via $\zeta(c)=2M/(n-1)$, we may invoke Lemma~4.2 with $q_0=M$ and $\sigma^2(c,n)\asymp k_0=M-n=m$. Thus,
\[
\mathbf{P}\!\left(E_p=m \,\middle|\, S_k\!\ge 0,\ k<n\right)
=\mathbf{P}\!\left(X_p(n)=M\right)
=\frac{1+o(1)}{\sqrt{2\pi}\,\sigma(c,n)}.
\]
Consequently, the expected number of restarts is $O(\sqrt{m})$, so with $O(n)$ work per trial the expected cost of Step~1 is $O(n\sqrt{m})$.

\section{Conclusion}\label{conclusion}

In this work, we have proposed an exact method for generating connected $G(n,p)$ graphs, based on step-by-step vertex exploration (constructing an exploration tree) followed by adding the remaining edges with the original probability $p$. The key observation is the correspondence between exploration trajectories and the conditional distribution of Poisson (or multinomial) random walks; this allows us to apply rejection sampling only when generating the exploration steps, thereby avoiding an inefficient search over the entire graph space.

Our analysis shows that the proposed algorithm samples from the correct distribution (i.e., the $G(n,p)$ model conditioned on connectivity). In the sparse regime $p=c/n$ its \emph{expected} running time is $O(n)$, while for arbitrary $p$ the worst-case cost is $O(n^2)$. 

Moreover, building on this algorithm, one can generate connected graphs in the $G(n,M)$ model by an additional acceptance–rejection step on the number of edges: Steps~2–3 have the same costs as in the $G(n,p)$ case, and for Step~1 the expected cost scales as $O(n\sqrt{m})$ with $m=M-(n-1)$.

A promising direction for further research is to extend the proposed approach to other classes of random graphs. For example, one can similarly develop a scheme for generating connected bipartite graphs (the models $G(n_1, n_2, p)$ and $G(n_1, n_2, M$) by using an appropriate modification of the Poisson random walk and exploration tree for the bipartite case. It is expected that the resulting methods will generate connected bipartite graphs according to the desired (conditional) distribution in the sparse regime taking a polynomial time.

\paragraph{ACKNOWLEDGMENTS}
\noindent The author is sincerely grateful to A.\,V.~Shklyaev for guidance and to M.\,M.~Koshelev for his insightful comments and advices.

\paragraph{Code Availability}
An open-source implementation of the proposed generators is available at \url{https://github.com/bchinyaev/connected_erdos_renyi}.


\end{document}